\documentclass[prc,superscriptaddress,unsortedaddress,twocolumn,showpacs]{revtex4}
\usepackage{amsmath}
\usepackage{amssymb}
\usepackage{times}
\usepackage{mathrsfs}
\usepackage{multirow}
\usepackage{bm}
\usepackage{wrapft}
\usepackage{color}
\usepackage{graphicx}
\usepackage{ulem}

\def\fun#1#2{\lower3.6pt\vbox{\baselineskip0pt\lineskip.9pt
\ialign{$\mathsurround=0pt#1\hfil##\hfil$\crcr#2\crcr\sim\crcr}}}

\newcommand{\beq}{\begin{equation}}
\newcommand{\eeq}{\end{equation}}
\newcommand{\bea}{\begin{eqnarray}}
\newcommand{\eea}{\end{eqnarray}}

\newcommand{\bfi}[1]{\mbox{\boldmath $#1$}}

\newcommand{\vp}{{\bfi p}}

\newcommand{\vs}{{\bfi s}}

\newcommand{\vt}{{\bfi t}}

\renewcommand{\vr}{{\bfi r}}
\newcommand{\vR}{{\bfi R}}

\begin{document}

\title{
Consistency between the monopole strength of the Hoyle state determined
by structural calculation and that extracted from reaction observables
}

\author{Kosho Minomo}
\email[]{minomo@rcnp.osaka-u.ac.jp}
\affiliation{Research Center for Nuclear Physics, Osaka University, Ibaraki 567-0047, Japan}

\author{Kazuyuki Ogata}
\affiliation{Research Center for Nuclear Physics, Osaka University, Ibaraki 567-0047, Japan}

\date{\today}

\begin{abstract}
We analyze the $\alpha$-$^{12}$C inelastic scattering to the $0_2^+$ state of $^{12}$C, the Hoyle state,
in a fully microscopic framework.
With no free adjustable parameter, the inelastic cross sections at forward angles are
well reproduced by the microscopic reaction calculation using
the transition density of $^{12}$C obtained by the resonating group method
and the nucleon-nucleon $g$ matrix interaction developed by the Melbourne group.
It is thus shown that the monopole transition strength obtained
by the structural calculation is consistent with that extracted from
the reaction observable, suggesting no missing monopole strength of the Hoyle state.
\end{abstract}


\maketitle

The $0_2^+$ state of $^{12}$C, the so-called Hoyle state, has intensively been studied
theoretically and experimentally~\cite{Ueg77,Kam81,Des87,Fun03,Ohk04,Che07,Eny07,Fre09,Rot11,Epe12,Fuk13,Zim13,Suh15}.
Despite a rather clear understanding of its three-$\alpha$ structure, the description of
the Hoyle state appeared in reaction observables, the ($\alpha,\alpha'$)
inelastic scattering cross section in particular, has not been achieved.
It was reported in many studies~\cite{Tak06,Kho08,Ito11,Kaw14} that
the ($\alpha,\alpha'$) cross section theoretically obtained with using
the transition density of $^{12}$C from the ground state to the $0_2^+$ state
significantly overshot the observed cross section. This puzzle is called
{\it the missing monopole strength of the Hoyle state}~\cite{Kho08}.

In these preceding studies, however, a semi-microscopic treatment
of the distorting potential between $\alpha$ and $^{12}$C
as well as the coupling potentials for the excitation of $^{12}$C
was adopted. This suggests some ambiguities in the distorting and
coupling potentials that connect the structural information and the
reaction observable.
In the present study we apply a fully microscopic framework to the
($\alpha,\alpha'$) inelastic scattering to the $0_2^+$ state of $^{12}$C,
and show that the calculated result agrees with the experimental cross
section and essentially there is no room for the missing monopole strength.

In this study we adopt the $g$-matrix folding model with the target-density
approximation (TDA)~\cite{Ega14,Toy15a,Toy15b}; the local density
of the target nucleus is used as an input density for the $g$ matrix.
This TDA $g$-matrix approach has been derived from the nucleus-nucleus
multiple scattering theory~\cite{Yah08} in Ref.~\cite{Toy15a} and shown
to work well for describing the elastic scattering of $^3$He~\cite{Toy15a}
and $\alpha$~\cite{Ega14,Toy15b} off several mid-heavy and heavy target
nuclei. We do not include the chiral three-nucleon force
modification to the $g$ matrix because its effect on $\alpha$ scattering
was shown to be very small~\cite{Toy15b}.

\begin{figure}[tbp]
\begin{center}
\includegraphics[width=0.35\textwidth,clip]{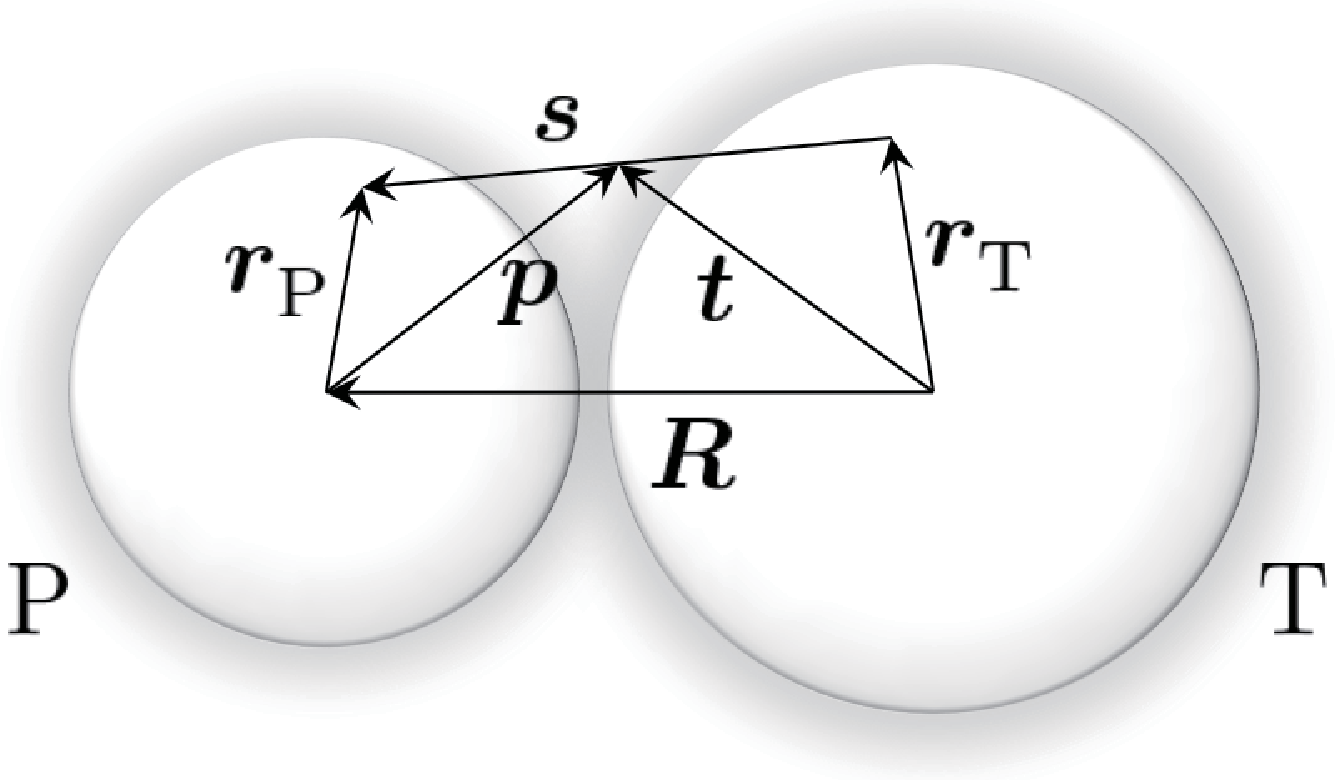}
\caption{(Color online)
Definition of the coordinates.
}
\label{fig1}
\end{center}
\end{figure}

We consider $\alpha$ as a projectile (P) and $^{12}$C as a target nucleus (T).
The $\alpha$ particle is assumed to stay in the ground state, whereas the transition of $^{12}$C
between the $0_1^+$, $2_1^+$, $0_2^+$, $3_1^-$, $2_2^+$, and $0_3^+$ states are explicitly taken into account.
The coupled-channel (CC) equation to be solved is given by
\begin{widetext}
\bea
&&
\bigg[-\frac{\hbar_{}^2}{2\mu}\frac{d^2}{dR^2}
+\frac{\hbar_{}^2}{2\mu}\frac{L(L+1)}{R_{}^2}
+F_{\gamma L,\gamma L}^{J}(R)
+U^{\rm Coul}(R)-E_\gamma^{}\bigg]
\chi_{\gamma L}^{J}(R)
=
-\sum_{\gamma'L'\neq \gamma L}^{}F_{\gamma L,\gamma'L'}^{J}(R)
\chi_{\gamma'L'}^{J}(R),
\nonumber\\
\label{cceq}
\eea
\bea
F_{\gamma L,\gamma'L'}^{J}(R)
&=&
\frac{1}{4\pi}\sum_{\lambda}
i^{L'-L+\lambda}
(-1)^{L'-L+I-J+\lambda}
\sqrt{(2L+1)(2L'+1)(2I+1)(2\lambda+1)}
W(LIL'I'|J\lambda)(L'0L0|\lambda 0)
\nonumber\\
&&\times
\bigg\{
\int\!d\hat{\vR}~d\vr_{\rm P}^{}~d\vr_{\rm T}^{}~
\rho_{\rm P}^{}(r_{\rm P}^{})
\rho_{\gamma\gamma'}^{\lambda}(r_{\rm T}^{})
g^{\rm (dr)}(s;\rho)
\big[Y_{\lambda}^{}(\hat{\vR})
\otimes
Y_{\lambda}^{}(\hat{\vr}_{\rm T}^{})
\big]_{00}^{}
\nonumber\\
&&+
\int\!d\hat{\vR}~d\vp~d\vs~
\rho_{\rm P}^{}(r_{\rm P}^{})
\hat{j}_1^{}\big(k_{\rm F}^{}(p)s\big)
\rho_{\gamma\gamma'}^{\lambda}(r_{\rm T}^{})
\hat{j}_1^{}\big(k_{\rm F}^{}(t)s\big)
g^{\rm (ex)}(s;\rho)
j_0^{}\big(k(R)s/M\big)
\big[Y_{\lambda}^{}(\hat{\vR})
\otimes
Y_{\lambda}^{}(\hat{\vt})
\big]_{00}^{}
\bigg\},
\nonumber\\
\label{couppot}
\eea
\end{widetext}
where $\chi_{\gamma L}^{J}(R)$ is the radial part of the P-T scattering
wave function in the $(\gamma L)$ channel; $\gamma$ specifies the
state of T and $L$ is the orbital angular momentum between P and T.
The total spin of T in the $\gamma$ state is denoted by $I$.
The definition of the coordinates is given in Fig.~\ref{fig1}.
%
%
$E_\gamma^{}$ is defined by $E_\gamma^{}=E-\varepsilon_\gamma^{}$
with $E$ the incident energy of P in the center-of-mass system and
$\varepsilon_\gamma^{}$ the the excitation energy of T.
$M$ is defined by $M=A_{\rm P}A_{\rm T}/(A_{\rm P}+A_{\rm T})$,
where $A_{\rm P(T)}$ is the mass number of P (T).
The Coulomb potential between P and T is denoted by $U^{\rm Coul}(R)$.
$Y_\lambda$ is the spherical harmonics, $j_n$ is the spherical
Bessel function, $(L'0L0\lambda 0)$ is the Clebsch-Gordan coefficient, and
$W(LSL'S'|J\lambda)$ is the Racah coefficient.

This coupled-channel approach with a $g$ matrix has widely been adopted
so far~\cite{Kho00,Sak88,Kat02,Min16}; the possible double-counting
for the coupling to non-elastic channels are expected to be negligible
as discussed in Ref.~\cite{Min16}.
Equation~(\ref{couppot}) contains two key ingredients. One is the
nuclear transition density
$\rho_{\gamma\gamma'}^{\lambda}$
and the other is the $g$ matrix $g^{\rm (dr/ex)}$; the superscript
dr (ex) indicates the direct (exchange) part of $g$, an explicit form
of which is shown in, e.g., Ref.~\cite{Bri77}.

We adopt the transition density of $^{12}$C obtained by the resonating
group method (RGM) based on a three-$\alpha$ model~\cite{Kam81};
the $0_1^+$, $2_1^+$ (4.44~MeV), $0_2^+$ (7.65~MeV), $3_1^-$ (9.64~MeV),
$2_2^+$ (9.84~MeV)~\cite{Ito11}, and $0_3^+$ (10.3~MeV) states are considered as mentioned above.
These densities are shown to reproduce the elastic and inelastic form factors
for electron scattering and are thus highly reliable.
In the CC calculation of the ($\alpha,\alpha'$) process, we include
all the six states listed above; we use the experimental values
of the excitation energies.
For the ground state density of $\alpha$, $\rho_{\rm P}^{}$, we take the phenomenological one
determined from electron scattering~\cite{Vri87} in which the finite-size effect of
proton charge is unfolded with a standard procedure~\cite{Sin78}.

As for $g$, we use the Melbourne $g$-matrix interaction~\cite{Amo00},
which has been highly successful, with no free parameters, in
describing various nucleon-nucleus elastic and inelastic cross sections in
a wide range of incident energies.
The use of a $g$-matrix interaction having a predictive power
is one of the most essential features of the present study.
As mentioned, we use the TDA for evaluating the argument $\rho$ in
$g^{\rm (dr/ex)}_{}$, i.e.,
$\rho=\rho_{\gamma\gamma}^{\lambda=0}(r_{\rm m}^{})$, where
$r_{\rm m}^{}$ denotes the midpoint of the interacting two nucleons.
For the nondiagonal potentials, we take the average of the densities
in the initial and final states, i.e.,
$\rho_{\rm T}^{}=[\rho_{\gamma\gamma}^{\lambda=0}(r_{\rm m}^{})+\rho_{\gamma'\gamma'}^{\lambda=0}(r_{\rm m}^{})]/2$.

\begin{figure}[tbp]
\begin{center}
\includegraphics[width=0.45\textwidth,clip]{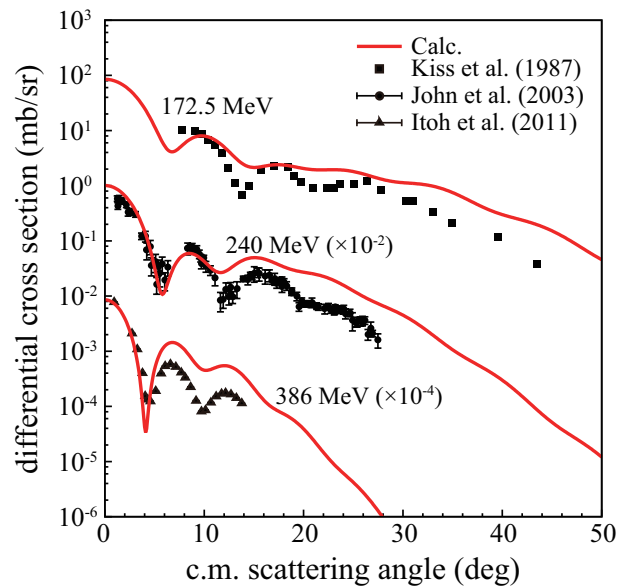}
\caption{(Color online)
Differential cross sections of $\alpha$-$^{12}$C inelastic scattering
to the $0_2^+$ state at 172.5, 240, and 386 MeV, as a function of the
scattering angle in the center-of-mass system.
The experimental data are taken from Refs.~\cite{Kis87,Joh03,Ito11}.
}
\label{fig2}
\end{center}
\end{figure}

\begin{figure}[tbp]
\begin{center}
\includegraphics[width=0.45\textwidth,clip]{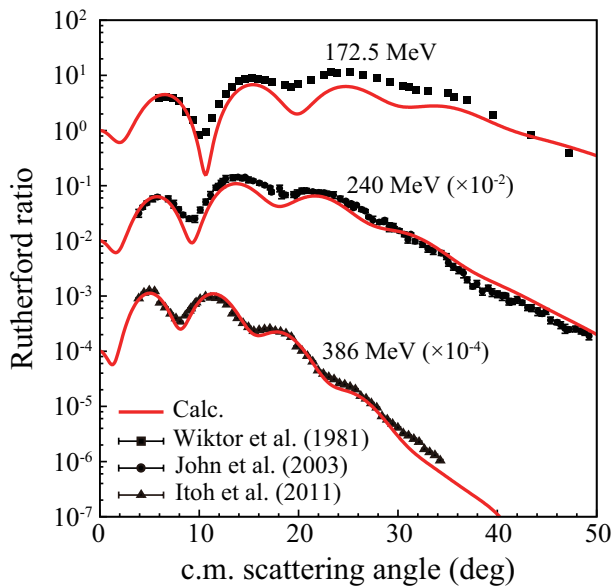}
\caption{(Color online)
Same as in Fig.~\ref{fig2} but for the elastic scattering
(the ratio to the Rutherford cross section).
The experimental data are taken from Refs.~\cite{Wik81,Joh03,Ito11}.
}
\label{fig3}
\end{center}
\end{figure}

Figure~\ref{fig2} shows the differential cross sections of
$^{12}$C($\alpha,\alpha'$)$^{12}$C($0_2^+$)
at 172.5, 240, and 386~MeV, compared with the experimental
data~\cite{Kis87,Joh03,Ito11}.
%
%
One sees that with no free adjustable parameters the inelastic
cross section to the $0_2^+$ state of $^{12}$C is reproduced well
at forward angles at these three energies. At larger angles
the calculation slightly overshoots the experimental data.
However, there seems to be no room for the missing monopole
strength; if it could exist, the inelastic cross section would
decrease at all angles and the good agreement at forward angles
would be lost, though it is not so clear at 172.5~MeV
because of the lack of data at very forward angles.
We show the results for the elastic scattering in Fig.~\ref{fig3}
in the same way as in Fig.~\ref{fig2}; a very good agreement
between the calculated cross section and the experimental data
is obtained, which confirms the reliability of the microscopic
reaction calculation adopted.
%

In a recent work~\cite{Cuo13} it was shown that the monopole transition
strength, $4.5\pm0.5 \; e~{\rm fm}^2$, to the $0_2^+$ state determined by comparing
the theoretical result with the experimental data is consistent with
that deduced from electron scattering. However, the antisymmetrized
molecular dynamics calculation adopted in the reaction analysis
gives a larger value (6.6~$e~{\rm fm}^2$) of the transition strength. Therefore
in Ref.~\cite{Cuo13} the structural input and the reaction observables
still have a gap of about 30\%. Furthermore, the interaction strength
was adjusted so as to reproduce the elastic scattering data; the
renormalization factor for the real (imaginary) part is 1.05 (1.27)
and 1.24 (1.38) for the analysis at 240 and 386~MeV, respectively.

\begin{figure}[tbp]
\begin{center}
\includegraphics[width=0.45\textwidth,clip]{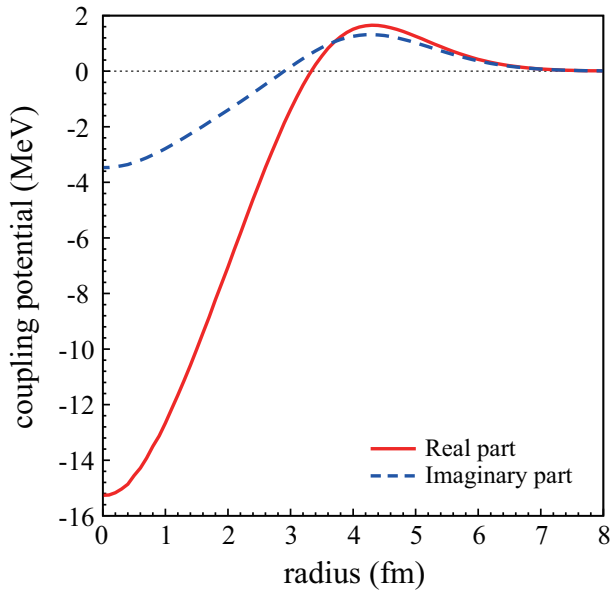}
\caption{(Color online)
Coupling potential between the $0_1^+$ and $0_2^+$ states of $^{12}$C
for the ($\alpha,\alpha'$) scattering at 172.5~MeV.
The solid (dashed) line corresponds to the real (imaginary) part of
the coupling potential.
}
\label{fig4}
\end{center}
\end{figure}

Finally, we show by the solid (dashed) line in Fig.~\ref{fig4}
the real (imaginary) part of the
coupling potential between the $0_1^+$ and $0_2^+$ states of
$^{12}$C for the $\alpha$ inelastic scattering at 172.5~MeV.
%
%
An important characteristic of the coupling potential is that it has a
peak at the origin and well concentrated in the nuclear interior region.
In Ref.~\cite{Kaw14} it was shown phenomenologically that this behavior
of the coupling potential is essential to reproduce the absolute value of the
$^{12}$C($\alpha,\alpha'$)$^{12}$C($0_2^+$) cross section. It should be
noted, however, that the origin of this behavior in the present study
is completely different from that in Ref.~\cite{Kaw14}.

In summary, we have calculated the $\alpha$-$^{12}$C inelastic cross section to the $0_2^+$ of $^{12}$C
with a microscopic coupled-channels method using the RGM transition density of $^{12}$C and
the Melbourne $g$-matrix interaction.
We have obtained a good agreement between the calculated and measured values
of the inelastic cross section at forward angles, as well as that of the elastic cross section.
It suggests that the monopole transition strength obtained by
the RGM calculation is consistent with the value extracted from the reaction observable.
Thus, it is concluded that there is no missing monopole strength of the Hoyle state.

The authors thank M.~Kamimura for providing the $^{12}$C RGM densities.
We also thank Y.~Kanada-En'yo, Y.~Funaki, T.~Kawabata, and S.~Adachi
for their interest in this work and for useful discussions.
The numerical calculations in this work were performed at RCNP.
This work is supported in part by
Grant-in-Aid for Scientific Research (No. 25400255)
from the Japan Society for the Promotion of Science (JSPS)
and by the ImPACT Program of Council for Science,
Technology and Innovation (Cabinet Office, Government of Japan).



\begin{thebibliography}{00}

\bibitem{Ueg77}
E. Uegaki, S. Okabe, Y. Abe, and H. Tanaka,
Prog. Theor. Phys. {\bf 57}, 1262 (1977).

\bibitem{Kam81}
M. Kamimura,
Nucl. Phys. A {\bf 351}, 456 (1981).

\bibitem{Des87}
P. Descouvemont and D. Baye,
Phys. Rev. C {\bf 36}, 54 (1987).

\bibitem{Fun03}
Y. Funaki, A. Tohsaki, H. Horiuchi, P. Schuck, and G. R{\" o}pke,
Phys. Rev. C {\bf 67}, 051306 (2003).

\bibitem{Ohk04}
S. Ohkubo and Y. Hirabayashi,
Phys. Rev. C {\bf 70}, 041602 (2004).

\bibitem{Che07}
M. Chernykh, H. Feldmeier, T. Neff, P. von Neumann-Cosel, and A. Richter,
Phys. Rev. Lett. {\bf 98}, 032501 (2007).

\bibitem{Eny07}
Y. Kanada-En'yo,
Prog. Theor. Phys. {\bf 117}, 655 (2007).

\bibitem{Rot11}
R. Roth, J. Langhammer, A. Calci, S. Binder, and P. Navr{\' a}til,
Phys. Rev. Lett. {\bf 107}, 072501 (2011).

\bibitem{Epe12}
E. Epelbaum, H. Krebs, T. A. L{\" a}hde, D. Lee, and Ulf-G. Mei{\ss}ner,
Phys. Rev. Lett. {\bf 109}, 252501 (2012).

\bibitem{Fre09}
M. Freer {\it et al.},
Phys. Rev. C {\bf 80}, 041303 (2009).

\bibitem{Fuk13}
Y. Fukuoka, S. Shinohara, Y. Funaki, T. Nakatsukasa, and K. Yabana,
Phys. Rev. C {\bf 88}, 014321 (2013).

\bibitem{Zim13}
W. R. Zimmerman {\it et al.},
Phys. Rev. Lett. {\bf 110}, 152502 (2013).

\bibitem{Suh15}
T. Suhara and Y. Kanada-En'yo,
Phys. Rev. C {\bf 91}, 024315 (2015).

\bibitem{Tak06}
M. Takashina and Y. Sakuragi,
Phys. Rev. C {\bf 74}, 054606 (2006).

\bibitem{Kho08}
D. T. Khoa and D. C. Cuong,
Phys. Lett. B {\bf 660}, 331 (2008).

\bibitem{Ito11}
M. Itoh {\it et al}.,
Phys. Rev. C {\bf 84}, 054308 (2011).

\bibitem{Kaw14}
T. Kawabata \textit{et al}.,
J. Phys.: Conf. Series {\bf 569}, 012014 (2014).

\bibitem{Ega14}
K. Egashira, K. Minomo, M. Toyokawa, T. Matsumoto, and M. Yahiro,
Phys. Rev. C {\bf 89}, 064611 (2014).

\bibitem{Toy15a}
M. Toyokawa, T. Matsumoto, K. Minomo, and M. Yahiro,
Phys. Rev. C {\bf 91}, 064610 (2015).

\bibitem{Toy15b}
M. Toyokawa, M. Yahiro, T. Matsumoto, K. Minomo, K. Ogata, and M. Kohno,
Phys. Rev. C {\bf 92}, 024618 (2015).

\bibitem{Yah08}
M. Yahiro, K. Minomo, K. Ogata, and M. Kawai,
Prog. Theor. Phys. {\bf 120}, 767 (2008).

\bibitem{Sak88}
Y. Sakuragi, M. Yahiro, M. Kamimura, M. Tanifuji,
Nucl. Phys. A {\bf 480}, 361 (1988).

\bibitem{Kho00}
D. T. Khoa and G. R. Satchler,
Nucl. Phys. A {\bf 668}, 3 (2000).

\bibitem{Kat02}
M. Katsuma, Y. Sakuragi, S. Okabe, and Y. Kond$\bar{\rm o}$,
Prog. Theor. Phys. {\bf 107}, 377 (2002).

\bibitem{Min16}
K. Minomo, M. Kohno, and K. Ogata,
Phys. Rev. C {\bf 93}, 014607 (2016).

\bibitem{Bri77}
F. A. Brieva and J. R. Rook, Nucl. Phys. A {\bf 291}, 299 (1977);
{\it ibid.} 291, 317 (1977); {\it ibid.} 297, 206 (1978).

\bibitem{Vri87}
H. de Vries, C. W. de Jager, and C. de Vries,
At. Data Nucl. Data Tables {\bf 36}, 495 (1987).

\bibitem{Sin78}
R. P. Singhal, M. W. S. Macauley, and P. K. A. De Witt Huberts,
Nucl. Instrum. and Method {\bf 148}, 113 (1978).

\bibitem{Amo00}
K. Amos, P. J. Dortmans, H. V. von Geramb, S. Karataglidis, and J. Raynal,
Adv. Nucl. Phys. {\bf 25}, 275 (2000).

\bibitem{Joh03}
B. John, Y. Tokimoto, Y. W. Lui, H. L. Clark, X. Chen, and D. H. Youngblood,
Phys. Rev. C {\bf 68}, 014305 (2003).

\bibitem{Wik81}
S. Wiktor, C. M. B{\" o}ricke, A. Kiss, M. Rogge, P. Turek, and H. Dabrowski,
Acta Phys. Pol. B {\bf 12}, 491 (1981).

\bibitem{Kis87}
A. Kiss, C. M. B{\" o}ricke, M. Rogge, P. Turek, and S. Wiktor,
J. Phys. G: Nucl. Phys. {\bf 13}, 1067 (1987).

\bibitem{Cuo13}
D. C. Cuong, D. T. Khoa, and Y. Kanada-En'yo,
Phys. Rev. C {\bf 88}, 064317 (2013).




\end{thebibliography}
\end{document}